\documentclass[twocolumn,pra,showpacs,amsmath,amssymb]{revtex4}

\usepackage{amssymb,amsmath,amsfonts,amsthm,graphicx,color}

\begin{document}

\title{Maximal violation of the I3322 inequality using infinite dimensional quantum systems}
\author{K\'aroly F.~P\'al}
\email{kfpal@atomki.hu}
\author{Tam\'as V\'ertesi}
\email{tvertesi@dtp.atomki.hu}
\affiliation{Institute of Nuclear Research of the Hungarian Academy of Sciences\\
H-4001 Debrecen, P.O.~Box 51, Hungary}

\def\CC{\mathbb{C}}
\def\RR{\mathbb{R}}
\def\one{\leavevmode\hbox{\small1\normalsize\kern-.33em1}}
\newcommand*{\tr}{\mathsf{Tr}}
\newcommand{\diag}{\mathop{\mathrm{diag}}}
\newcommand{\ket}[1]{|#1\rangle}
\newcommand{\bra}[1]{\langle#1|}
\newtheorem{theorem}{Theorem}
\newtheorem{lemma}{Lemma}

\date{\today}

\begin{abstract}
The $I_{3322}$ inequality is the simplest bipartite two-outcome Bell inequality beyond the
Clauser-Horne-Shimony-Holt (CHSH) inequality, consisting of three two-outcome measurements per party. 
In case of the CHSH inequality the maximal quantum violation can already be attained with local two-dimensional quantum systems, however, there is no such evidence for the $I_{3322}$ inequality. In this paper a family of measurement operators and states is given which enables us to attain the largest possible quantum value in an infinite dimensional Hilbert space. Further, it is conjectured that our construction is optimal in the sense that measuring finite dimensional quantum systems is not enough to achieve the true quantum maximum. We also describe an efficient iterative algorithm for computing quantum maximum of an arbitrary two-outcome Bell inequality in any given Hilbert space dimension.  This algorithm played a key role to obtain our results for the $I_{3322}$ inequality, and we also applied it to improve on our previous results concerning the maximum quantum violation of several bipartite two-outcome Bell inequalities with up to five settings per party.
\end{abstract}

\pacs{03.65.Ud, 03.67.-a} 
\maketitle

\section{Introduction}\label{intro}
One of the most puzzling features of quantum theory is its nonlocal nature. Separated observers on a shared entangled state may carry out measurements on such a way that the correlations they generate are outside the set of common cause correlations \cite{Bell}. In particular, such quantum correlations may find application in novel device-independent information tasks, which have no counterparts in the classical world. They enable perfect security \cite{ANGMPS}, randomness generation \cite{Pironio} and state tomography \cite{Bardyn} without the need to trust the internal working of the devices.

The concept of Bell inequalities is a particularly useful tool to detect nonlocal quantum correlations, as violation of a single 
Bell inequality conclusively proves the nonlocal character of correlations. The standard scenario for a bipartite two-outcome Bell test is as follows. Two spacelike separated parties, Alice and Bob, both share copies of a quantum state $|\psi\rangle$ of a given dimension $n\times n$. Alice (Bob) may choose between $m_A$ ($m_B$) alternative measurements at random, where each measurement has two possible outcomes $\{0,1\}$. In a single run of the experiment the correlations between the two $\{0,1\}$-valued observables $A_i$ and $B_j$ can be represented by the product $A_iB_j$. In order to obtain an accurate estimation of the correlations for each pair $(i,j)$, Alice and Bob repeat the experiment many times using a copy of the state $|\psi\rangle$ in each round. Averaging over many runs of the experiment yields the mean value $\langle A_iB_j\rangle$. 

The CHSH inequality \cite{CHSH} is probably the most well-known and simplest example of a Bell inequality consisting of two measurement settings ($m_A=m_B=2$) both on Alice's and on Bob's part. Any correlations in the framework of local classical theories obey the following CHSH inequality \cite{CHSH},
\begin{equation}
\langle A_1B_1\rangle+\langle A_1B_2\rangle+\langle A_2B_1\rangle-\langle A_2B_2\rangle
-\langle A_1\rangle -\langle B_1\rangle\le 0. \label{eq:CHSH}
\end{equation}
Curiously, quantum theory allows for a violation of the CHSH inequality, but the strength of nonlocal correlations
is still limited, it obeys \cite{Tsirelson}
\begin{align}
&\langle A_1B_1\rangle+\langle A_1B_2\rangle+\langle A_2B_1\rangle-\langle A_2B_2\rangle -\langle
A_1\rangle -\langle B_1\rangle\nonumber\\
&\le 1/\sqrt 2 -1/2. 
\label{eq:Tsirelson}
\end{align}
where now the expectation values can be expressed by $\langle
A_iB_j\rangle=\langle\psi|\hat A_i\otimes \hat B_j|\psi\rangle$, $\langle
A_i\rangle=\langle\psi|\hat A_i\otimes \hat I_B|\psi\rangle$, $\langle
B_j\rangle=\langle\psi|\hat I_A\otimes \hat B_j|\psi\rangle$, $i,j=1,2$ with
$\{0,1\}$-valued observables $\hat A_i$, $\hat B_j$. According to Tsirelson's
theorem \cite{Tsirelson}, the bound applies to any quantum
correlations without making assumptions on the sort of
measurements or the dimensionality of the states involved. Though,
the maximum value of $1/\sqrt 2 -1/2$ can already be achieved with a
maximally entangled two-qubit state.
%$|\psi^+\rangle=(|00\rangle+|11\rangle)/\sqrt2$. 

Inequalities, whose bound are saturated with quantum correlations without relying on dimensionality 
such as in Eq.~(\ref{eq:Tsirelson}) were coined as quantum Bell inequalities \cite{QBell}. There exist various methods in the literature \cite{Masanes,Wehner,NPA1}, to cite just a few, which enable one to derive quantum Bell inequalities. Also, several explicit constructions  exist in the literature (e.g.,  \cite{CHTW, NPA2, PV09, Zohren}), including so-called irrelevant ones \cite{irrelevant} (where the classical and quantum limits coincide). The method invented by Navascu\'es, Pironio and Ac\'in (NPA) \cite{NPA1}  is based on the solution of a hierarchy of semidefinite programming (SDP) relaxations and is particularly useful since it gives better and better upper bounds on the maximum violation of an arbitrary Bell inequality by stepping to higher levels in the hierarchy. Moreover, the series of upper bounds in the hierarchy keep to the exact quantum maximum in terms of commuting measurements \cite{NPA2,DLTW}. On the other hand, one can use heuristic algorithms to obtain nontrivial lower bounds in some finite dimensional Hilbert spaces on Bell inequalities, recovering the explicit form of the states and measurement operators as well. If the above computed upper and lower bounds coincide within numerical accuracy for a given Bell inequality, then we may say that a quantum Bell inequality has been obtained, which delimits the boundary of the quantum domain.

In particular, in our previous papers \cite{PV08,PV09} we computed lower bounds on the maximum quantum violations and determined the corresponding
measurement operators and state vectors with two different methods, both involved some parametrization of the operators and applying a downhill simplex method to find the optimum parameters. The disadvantage of these methods is that the size of the Hilbert
space was very limited. We could actually handle systems of maximum eight dimensional real or six dimensional complex component spaces. Different sizes of component spaces require different parametrizations for the operators, hence in order to extend these methods to a higher
dimension would involve working out an appropriate parametrization. Moreover, the choice of measurement operators were also limited. For example, in the case of eight dimensional component spaces only projection operators projecting onto four dimensional subspaces were allowed. 

In spite of the limitations we could get the maximum quantum violation of the great majority of inequalities we considered (the list comprises 241 bipartite Bell inequalities with up to five settings per party collected from Refs.~\cite{BG,Avis,PV09} and detailed results concerning their optimum violations are presented in the web page \cite{Bellweb}). However, there were still a few exceptions, where the upper bound value resulting from the NPA method \cite{NPA1} did not match the best lower bound result. The most interesting one was the case of $I_{3322}$. This is the smallest case we considered, and perhaps the simplest tight Bell inequality after the CHSH one, with only three measurement settings per party. It was introduced by Froissart \cite{Froissart} back in 1981, and recently reinvented in Refs.~\cite{Sliwa,I3322}. It reads
\begin{align}
I_{3322}&\equiv-\langle A_2\rangle-\langle B_1\rangle-2\langle
B_2\rangle\nonumber\\
&+\langle A_1B_1\rangle+\langle A_1B_2\rangle+\langle A_2B_1\rangle+\langle A_2B_2\rangle\nonumber\\
&-\langle A_1B_3\rangle+\langle A_2B_3\rangle-\langle A_3B_1\rangle+\langle A_3B_2\rangle\le 0.
\label{eq:I3322}
\end{align}
In a local classical model we have the maximum value of 0, while the largest violation one could get with qubits was 0.25, which could already be achieved with a maximally entangled pair of qubits (see e.g.,~\cite{I3322,BG,NPA2,DLTW,PV08,PV09}). On the other hand, the best upper bounds are based on the NPA method \cite{NPA1} and at level three it yields the significantly higher upper bound, $0.250~875~56$ \cite{NPA2,DLTW}. We could even go above level three to an intermediate level in \cite{PV09}, and presently we have got the upper bound $0.250~875~38$ at level four. From the dependence of the bound on the level it was derived it seemed to be clear that it would not go much lower (note also that the computational complexity of the SDP problem increases dramatically with higher levels of relaxations). In fact, it is the above observation, which motivated us to search for a quantum violation beyond the two-qubit value of $0.25$.

In the present paper we set out to resolve the puzzling problem concerning the maximum quantum violation of the $I_{3322}$ inequality. To this end, we introduce in Sec.~\ref{ital} an efficient iterative algorithm which was applied to explore the largest quantum violation of $I_{3322}$ for states up to local dimension 20. Sec.~\ref{const} contains the main results, presenting the explicit construction of states and measurement operators, which give by means of an iterative method the conjectured maximal quantum value of $I_{3322}$ in function of the dimensionality. In particular, for dimensions very large we recover the upper bound computed with the NPA method at level four within high numerical accuracy, thereby establishing a quantum Bell inequality for $I_{3322}$. We note that the optimal quantum state in the infinite dimensional space is far from  the maximally entangled state, thereby supporting the claim that entanglement and nonlocality are different resources \cite{MS}. In Sec.~\ref{other} we investigate the remaining 19 Bell inequalities from a set of 241 inequalities (plus two additional symmetric 4-setting inequalities from Ref.~\cite{BGP}), where the tightness of the quantum bound could not be proven previously, and in some of the remaining cases now we manage to close the gap. Sec.~\ref{conc} summarizes the results achieved. 

\section{The iterative algorithm}\label{ital}

As mentioned above, the drawback of our previous heuristic methods in Refs.~\cite{PV08,PV09} for computing lower bounds
on the maximum quantum violation of Bell inequalities is that
the size of the local Hilbert space we could handle is limited up to dimension eight. 
With the iterative method we introduce here for two-outcome Bell inequalities we can go to higher dimensions, which is
only limited by the computational difficulties due to the increasing complexity
of the problem with increasing Hilbert space dimensions. Similarly to the previous
methods, this algorithm does not guarantee to converge to the global optimum solution.
In case of larger spaces
we may miss it even after tens of thousands of restarts with different initial values.
Nevertheless, it managed to find every optima we derived
with our previous methods, and in almost all cases it has done it considerably faster.

The problem to be solved consists of maximizing the quantum value of the Bell expression with coefficients $M_{\mu\nu}$,
which can be written as:
\begin{equation}
{\cal Q}={\max}\sum_{\mu=0}^{m_A}\sum_{\nu=0}^{m_B}M_{\mu\nu}\langle\psi|\hat A_{\mu}\otimes\hat B_{\nu}|
\psi\rangle,
\label{eq:bellexop}
\end{equation}
where there are $m_A$ and $m_B$ measurement settings for Alice and Bob, respectively,
$\hat A_\mu$ ($1\leq\mu\leq m_A$) and $\hat B_\nu$ ($1\leq\nu\leq m_B$) are the measurement
operators of Alice and Bob, respectively, $\hat A_0\equiv\hat I_A$ and $\hat B_0\equiv\hat I_B$ are unit
operators in the component spaces, and $|\psi\rangle$ is the state vector. The maximum can always
be reached by a pure state. In case of two-outcome measurements it is enough to consider
projection operators as the measurement operators \cite{CHTW}. When there are only two parties,
a further simplification is that when we take a matrix representation
of the expression above, we may choose the bases in the component spaces such that
they correspond to the Schmidt decomposition of the state vector $|\psi\rangle$. If we confine
ourselves to $n$-dimensional component spaces we get:
\begin{equation}
{\cal Q}_n={\max}\sum_{\mu=0}^{m_A}\sum_{\nu=0}^{m_B}\sum_{i=1}^n\sum_{j=1}^nM_{\mu\nu}
A^\mu_{ij}B^\nu_{ij}\lambda_i\lambda_j,
\label{eq:bellexmat}
\end{equation}
where $A^\mu_{ij}$ and $B^\nu_{ij}$ are components of matrices of $\hat A_\mu$ and
$\hat B_\nu$, respectively, and $\lambda_i$ are the Schmidt coefficients of the
state vector.
If $n$ is smaller than the dimensionality of the component spaces required for the maximum
quantum violation, then ${\cal Q}_n\leq{\cal Q}$.

To solve the problem first we choose appropriate random matrices and numbers for Bob's
measurement operators, and for $\lambda_i$. To get the matrices, first we take a diagonal
matrix with zero and one diagonal values chosen randomly, then we apply many random two
dimensional unitary (or if we confine ourselves to real matrices, orthogonal)
transformations. For $\lambda_i$ we take positive numbers between zero and one from a
uniform distribution, then we normalize them. In the iterative algorithm the first
step is to calculate the optimal measurement operators of Alice, given Bob's operators and
the state vector. This can be done directly. Equation (\ref{eq:bellexmat}) can be rewritten
as:
\begin{equation}
{\cal Q}_n={\max} \sum_{\mu=0}^{m_A}\sum_{i=1}^n\sum_{j=1}^nA^\mu_{ij}X^\mu_{ji}=
{\max}\sum_{\mu=0}^{m_A}{\rm Tr}(\hat A_{\mu}\hat X_{\mu}),
\label{eq:aopti}
\end{equation}
where
\begin{equation}
X^\mu_{ji}=\sum_{\nu=0}^{m_B}M_{\mu\nu}B^\nu_{ij}\lambda_i\lambda_j,
\label{eq:xmat}
\end{equation}
We can get the matrix of the optimum $\hat A_{\mu}$ ($1\leq\mu\leq m_A$) the following way.
First we diagonalize the matrix of $\hat X_{\mu}$. Then we create a diagonal matrix.
We choose its diagonal matrix element one where the diagonalized matrix of $\hat X_{\mu}$ contains
a positive number, and zero otherwise. Then we transform this matrix with the inverse of the
transformation
that diagonalized the matrix of $\hat X_{\mu}$. To show that this is the matrix of the
optimal $\hat A_{\mu}$ we note that trace is invariant to basis transformation, and it is
easy to see that the optimum matrix of projector $\hat A_{\mu}$ in the basis diagonalizing
$\hat X_{\mu}$ is the one we have chosen.
 
The second step is to derive Bob's optimal matrices while Alice's
matrices and the state vector are fixed. This can be done equivalently to the first step.
These two steps are the same as the steps of the see-saw algorithm
by Werner and Wolf \cite{WW}, who used it to get maximum violation with fixed (not necessarily pure) states. A variant of the see-saw algorithm for Bell inequalities with multiple outcomes was also devised~\cite{Ito,LD}.

In the third step the best state vector is calculated
for the measurement operators. It is done the same way as in our previous method \cite{PV09}. 
An algorithm applying this very third step has been also used by Liang et al.~\cite{LLD} to compute quantum optima for multiple-outcome Bell expressions. 
From Eq.~(\ref{eq:bellexmat}) it is clear that the optimal $\lambda_i$ corresponds to
the eigenvector belonging to the largest eigenvalue of the matrix
\begin{equation}
\sum_{\mu=0}^{m_A}\sum_{\nu=0}^{m_B}M_{\mu\nu}A^\mu_{ij}B^\nu_{ij},
\label{eq:bestvect}
\end{equation}
and the eigenvalue is just the value of the Bell expression. It is not sure that
the components of the largest eigenvector will all be positive real numbers.
If they are not, they can not be called Schmidt coefficients. However, this does
not affect anything, it is best just to leave them as they are. If we want
to see the final result for the state vector represented by its Schmidt
coefficients, it is enough to make the appropriate transformation at the end of the
calculation. We simply have to multiply rows and columns of either Alice's or
Bob's matrices with phase factors. 

The algorithm consists of repeating these three steps until the value of the Bell expression
reach convergency. 
As this number may only increase in each step, convergence is ensured.
What is not ensured is that we will converge to the global optimum of the problem.
Therefore, we have to repeat the full procedure many times with different initial values.
We may save computation time if we stop a run when it clearly goes towards an inferior
solution. For example, if we are interested in a solution belonging to a larger
Hilbert space, we may stop whenever too many components of the state vector falls
below some threshold.

It is not difficult to extend the method to more than two participants.
In that case there is no Schmidt decomposition, which would ensure that one can
choose the bases in the component spaces such that the $n^2$-dimensional state vector has only
$n$ nonzero components. However, one can still use Eq.~(\ref{eq:aopti}) to calculate
the optimum measurement operators for Alice, and analogous equations for the operators
of the other participants. The only difference is that the formula for the
matrix of $\hat X_{\mu}$ (and the corresponding matrices for the other participants)
will be somewhat more complicated than Eq.~(\ref{eq:xmat}), involving summations to more
indices. The optimum state vector can also be calculated as the eigenvector
belonging to the largest eigenvalue of a matrix, but this will be a more general matrix
in the full $n^2$-dimensional Hilbert space. As the third step of the algorithm in this
case involves the diagonalization of a much larger matrix than the first two steps,
it is more efficient to do it less frequently, that is to repeat the first and the second
steps several times before updating the state vector.

\section{Family of constructions}\label{const}

From the optimality of formula~(\ref{eq:bellexmat}) it follows that the matrices of the optimum $\hat A_1$, $\hat A_2$, $\hat A_3$, $\hat B_1$, $\hat B_2$
and $\hat B_3$ are such that they satisfy the following Eq.~(\ref{eq:bestA1}),
Eq.~(\ref{eq:bestA2}), Eq.~(\ref{eq:bestA3}), Eq.~(\ref{eq:bestB1}),
Eq.~(\ref{eq:bestB2}) and Eq.~(\ref{eq:bestB3}), respectively:
\begin{align}
&\sum_{i,j=1}^n A_{ij}^1(B_{ij}^1+B_{ij}^2-B_{ij}^3)\lambda_i\lambda_j &=\rm{max}\label{eq:bestA1}\\
&\sum_{i,j=1}^n A_{ij}^2(B_{ij}^1+B_{ij}^2+B_{ij}^3-\delta_{ij})\lambda_i\lambda_j&=\rm{max}\label{eq:bestA2}\\
&\sum_{i,j=1}^n A_{ij}^3(B_{ij}^2-B_{ij}^1)\lambda_i\lambda_j &=\rm{max}\label{eq:bestA3}\\
&\sum_{i,j=1}^n B_{ij}^1(A_{ij}^1+A_{ij}^2-A_{ij}^3-\delta_{ij})\lambda_i\lambda_j &=\rm{max}\label{eq:bestB1}\\
&\sum_{i,j=1}^n B_{ij}^2(A_{ij}^1+A_{ij}^2+A_{ij}^3-2\delta_{ij})\lambda_i\lambda_j&=\rm{max}\label{eq:bestB2}\\
&\sum_{i,j=1}^n B_{ij}^3(A_{ij}^2-A_{ij}^1)\lambda_i\lambda_j &=\rm{max}\label{eq:bestB3}
\end{align}
Meanwhile, $\lambda_i$ are the components of the eigenvector belonging to the maximum eigenvalue
of the matrix
\begin{align}
M_{ij}=&-(A_{ij}^2+B_{ij}^1+2B_{ij}^2)\delta_{ij}+(A_{ij}^1+A_{ij}^2)(B_{ij}^1+B_{ij}^2)\notag\\
&+A_{ij}^3(B_{ij}^2-B_{ij}^1)+B_{ij}^3(A_{ij}^2-A_{ij}^1).
\label{eq:bestlambda}
\end{align}
The matrices of the measurement operators we got with the iterative method described in Section~\ref{ital} as the best solutions with all $\lambda_i$ differing from zero had very special forms.
We found most of the matrix elements zero within expected numerical accuracy,
and for odd number of dimensions
it turned out that with some appropriate reordering of the bases we could
also make all matrices blockdiagonal with
one block having a single element, while the rest consisting of
two by two blocks, with all matrix elements real. In the cases of $\hat A_1$, $\hat A_2$ and
$\hat B_3$, the block of size one is the first one,
while in the cases of $\hat B_1$, $\hat B_2$ and $\hat A_3$ it is
the last one. The actual form of the matrix of
$\hat A_2$ may be written as
\begin{equation}
\hat A_2=\left(\begin{array}{cccccccc}
1&&&&&&&\\
&\frac{1-c_2}{2}&\frac{s_2}{2}&&&&&\\
&\frac{s_2}{2}&\frac{1+c_2}{2}&&&&&\\
&&&\frac{1-c_4}{2}&\frac{s_4}{2}&&&\\
&&&\frac{s_4}{2}&\frac{1+c_4}{2}&&&\\
&&&&&\ddots&&\\
&&&&&&\frac{1-c_{n-1}}{2}&\frac{s_{n-1}}{2}\\
&&&&&&\frac{s_{n-1}}{2}&\frac{1+c_{n-1}}{2}\\
\end{array}\right),
\label{eq:A2}
\end{equation}
with $s_i\equiv\sqrt{1-c_i^2}$ are all positive.
The matrix of $\hat A_1$ tuns out to be the same, but the offdiagonal elements
have a negative sign. The form of $\hat B_2$ is the following:
\begin{equation}
\hat B_2=\left(\begin{array}{cccccccc}
\frac{1+c_1}{2}&\frac{s_1}{2}&&&&&&\\
\frac{s_1}{2}&\frac{1-c_1}{2}&&&&&&\\
&&\frac{1+c_3}{2}&\frac{s_3}{2}&&&&\\
&&\frac{s_3}{2}&\frac{1-c_3}{2}&&&&\\
&&&&\ddots&&&\\
&&&&&\frac{1+c_{n-2}}{2}&\frac{s_{n-2}}{2}&\\
&&&&&\frac{s_{n-2}}{2}&\frac{1-c_{n-2}}{2}&\\
&&&&&&&1+c_n
\end{array}\right),
\label{eq:B2}
\end{equation}
The matrix element $B^2_{nn}$ may take the value of 1 and 0, when
$c_n=0$ and $c_n=-1$, respectively. For smaller dimensions we got the former value
for all best solutions, but it turned out that from dimensions larger than
$n=79$ the other value gives the better solution. The notation of
$B^2_{nn}\equiv 1+c_n$ will become clear later. The matrix of
$\hat B_1$ is the same as that of $\hat B_2$, but $B^1_{nn}=0$, and like
in the case of $\hat A_1$, the offdiagonal matrix elements are negative.
Matrices $\hat A_3$ and $\hat B_3$ have even simpler forms: $A^3_{nn}=B^3_{11}=1$,
while all elements of the two by two blocks are $1/2$. It is easy to check that
if all $\lambda_i>0$, then the matrices of $\hat A_3$ and $\hat B_3$ do satisfy
the optimality conditions Eq.~(\ref{eq:bestA3}) and Eq.~(\ref{eq:bestB3}), respectively.
The matrix of $\hat A_3$ has the same block structure as $\hat B_2-\hat B_1$, as
it should, and the optimality condition can be checked block by block.
The value of the single element block in the matrix of $\hat B_2-\hat B_1$, that is
$B^2_{nn}-B^1_{nn}$ is either 2 or 0, depending on $c_n$, in the first case
the optimum is $A^3_{nn}=1$, while in the other case the chosen value $1$ is just
as good as value $0$ would have been. The two by two blocks in the matrix of
$B^2_{nn}-B^1_{nn}$ have zero diagonal elements, and positive nondiagonal
elements. As all $\lambda_i$ have the same sign, the number they will be
multiplied with is also positive. Therefore, the optimum two by two block of the
matrix of $\hat A_3$ is the symmetric two-dimensional matrix with eigenvalues
$0$ and $1$ having the largest possible nondiagonal elements. This matrix is the
one whose all four elements are $1/2$.

From Eq.~(\ref{eq:bestA2}) we can get equations determining the parameters of the matrix of
$\hat A_2$  that is $c_i$ with $i$ even (see Eq.~(\ref{eq:A2})) in terms of the parameters of
the matrices of Bob's
measurement operators, that is $c_i$ with $i$ odd, and the Schmidt coefficients $\lambda_i$.
As the matrix of $\hat B_1+\hat B_2-\hat I$ is diagonal, the block structure of $\hat A_1$ is
the same as that of $\hat B_3$. The matrix of $\hat B_1+\hat B_2+\hat B_3-\hat I$ is:
\begin{align}
&\hat B_1+\hat B_2+\hat B_3-\hat I=\notag\\
&\left(\begin{array}{cccccccc}
c_1&&&&&&&\\
&\frac{1}{2}-c_1&\frac{1}{2}&&&&&\\
&\frac{1}{2}&\frac{1}{2}+c_3&&&&&\\
&&&\frac{1}{2}-c_3&\frac{1}{2}&&&\\
&&&\frac{1}{2}&\frac{1}{2}+c_5&&&\\
&&&&&\ddots&&\\
&&&&&&\frac{1}{2}-c_{n-2}&\frac{1}{2}\\
&&&&&&\frac{1}{2}&\frac{1}{2}+c_{n}\\
\end{array}\right),
\label{eq:B1B2B3I}
\end{align}

The value of $A^2_{11}$ is determined by the sign of
$c_1$, the $A^1_{11}=1$ value is correct if $c_1>0$, which
turns out to be true. Each $c_i$, where $i$ is even,
occurs only in one two by two block of the matrix of $\hat A_2$
(see Eq.~(\ref{eq:A2})).  Its optimum value depends on the
corresponding block in Eq.~(\ref{eq:B1B2B3I}), containing
$c_{i-1}$ and $c_{i+1}$. We denoted $B^2_{nn}$ as $1+c_n$ such that
the last block has the same form as the others.
Using Eq.~(\ref{eq:bestA2}) we
arrive at the equation determining the optimum value of $c_i$:
\begin{align}
&{(1-c_i)\over 2}\left({1\over 2}-c_{i-1}\right)\lambda_i^2-
{(1+c_i)\over 2}\left({1\over 2}+c_{i+1}\right)\lambda_{i+1}^2\notag\\
&+{s_i\over 2}\lambda_i\lambda_{i+1}
=\rm{max}
\label{eq:optci1}
\end{align}
Multiplying the equation by four, dropping terms constant
in $c_i$ and substituting $s_i=\sqrt{1-c_i^2}$ we get:
\begin{align}
&c_i[(1+2c_{i+1})\lambda_{i+1}^2-(1-2c_{i-1})\lambda_i^2]
+2\sqrt{1-c_i^2}\lambda_i\lambda_{i+1}\notag\\
&=\rm{max},
\label{eq:optci2}
\end{align}
from which it follows that
\begin{equation}
c_i={\tau_i\over\sqrt{\tau_i^2+4\lambda_i^2\lambda_{i+1}^2}},
\label{eq:optci3}
\end{equation}
where
\begin{equation}
\tau_i\equiv(1+2c_{i+1})\lambda_{i+1}^2-(1-2c_{i-1})\lambda_i^2.
\label{eq:optci4}
\end{equation}
With $c_i$ chosen this way Eq.~(\ref{eq:bestA1}), the optimality
condition for $\hat A_1$ is also satisfied. If we started from this
condition instead of the one for $\hat A_2$, we would have arrived
at exacly the same results.

\begin{center}
\begin{figure}
\vspace{0.0cm}
\includegraphics[width=\columnwidth]{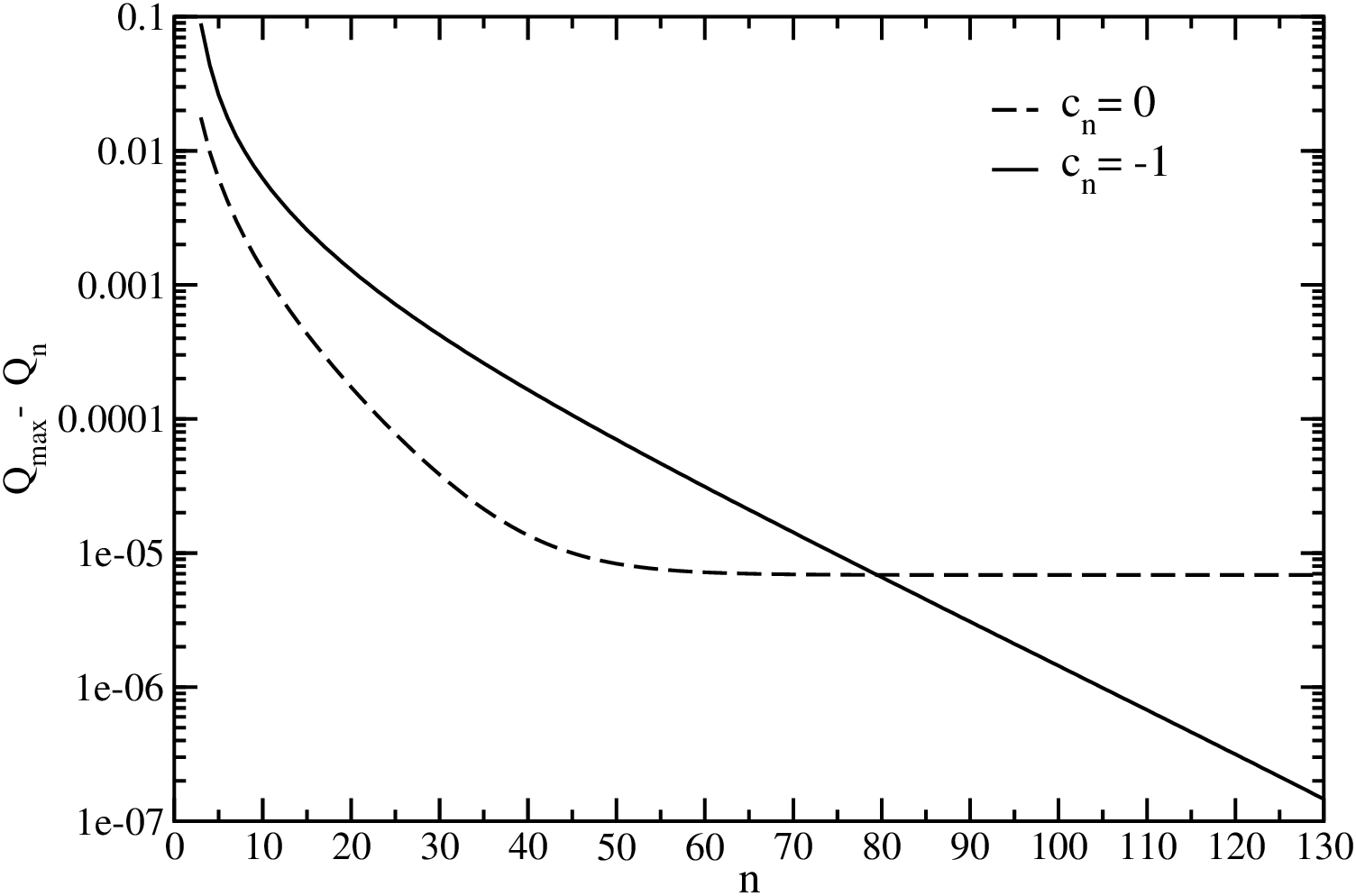} \caption{
Distance of ${\cal Q}_n$ from the upper bound of ${\cal Q}_{\rm max}=0.250~875~38$ is shown as function of the dimensionality for cases of $c_n=0$ and $c_n=-1$.} \label{convergence}
\end{figure}
\end{center}
 
In the same way as above, from Eq.~(\ref{eq:bestB1}) and Eq.~(\ref{eq:bestB2})
we can derive formulae for the optimum
values of the parameters of the matrices of $\hat B_1$ and
$\hat B_2$, that is for $c_i$, where $i$ is odd. These formulae
turn out to have exactly the same form as the ones derived
for even $i$ values, that is Eqs.~(\ref{eq:optci1}-\ref{eq:optci4}) are valid for
odd $i$. To get $c_1$, we have to introduce the notation $c_0\equiv 1$.
The value $B^1_{nn}=0$ is the optimum value.  Either $B^2_{nn}=0$
(that is $c_n=-1$) or $B^2_{nn}=1$ (that is $c_n=0$) is a possible choice.
Consistency with the optimality condition is fulfilled if $c_{n-1}\leq 0$ in
the first case, and $c_{n-1}\geq 0$ in the second one.

\begin{center}
\begin{figure}
\vspace{0.0cm}
\includegraphics[width=\columnwidth]{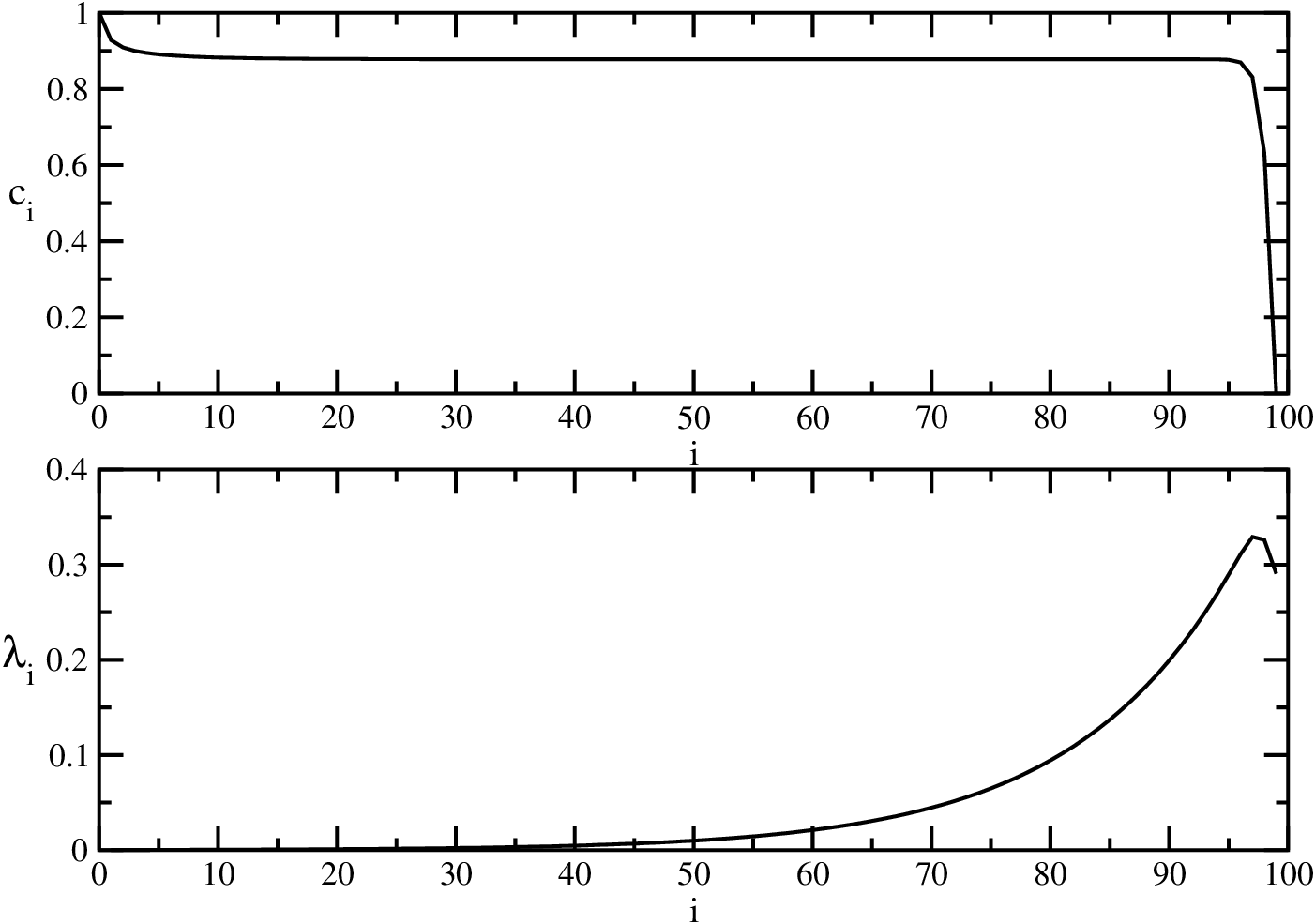} \caption{The values of $\lambda_i$ and $c_i$ are shown for $n=99$ with $c_n=0$.
} \label{aszimcl}
\end{figure}
\end{center}

\begin{center}
\begin{figure}
\vspace{0.0cm}
\includegraphics[width=\columnwidth]{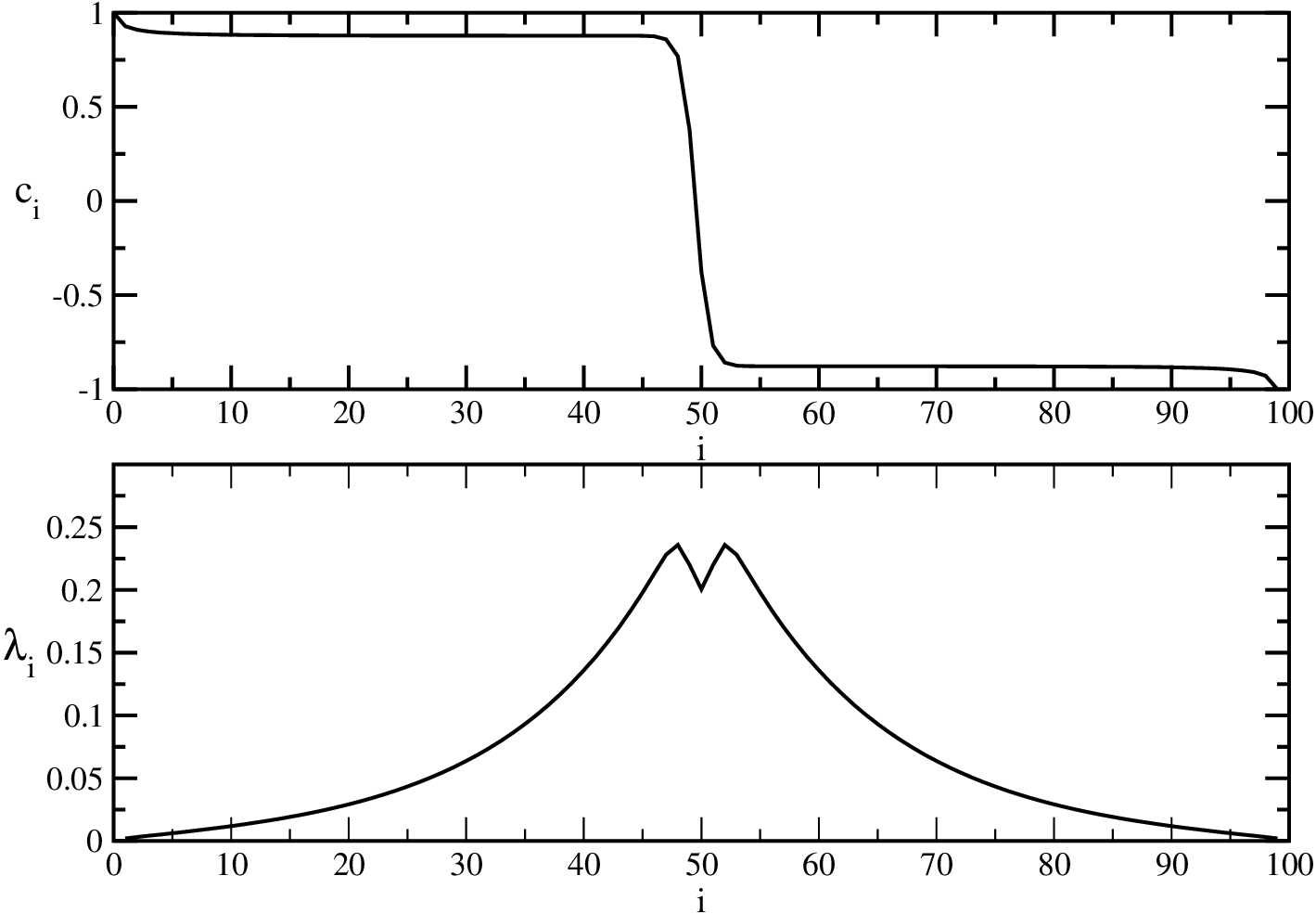} \caption{
The values of $\lambda_i$ and $c_i$ are shown for $n=99$ with $c_n=-1$.} \label{szimcl}
\end{figure}
\end{center}

With the specific forms of the measurement operators above, the matrix
$\hat M$ of Eq.~(\ref{eq:bestlambda}),
whose eigenvector corresponding to its largest eigenvalue
gives the Schmidt coefficients $\lambda_i$, is tridiagonal, and its nonzero
matrix elements are:
\begin{align}
&M_{ii}=c_{i-1}c_i+{{c_{i-1}-c_i}\over 2}-1+{{c_n+1}\over 2}\delta_{in},\notag\\
&M_{i(i+1)}={s_i\over 2}.
\label{eq:Mtridiag}
\end{align}
We may apply the iterative method with confining ourselves to
the present specific forms of the operators, and get solutions for
very large dimensions. This way we may increase the dimensionality
of the component Hilbert spaces even to a thousand, or more.
The distance of the maximum value of the Bell expression from
the upper bound $0.250~875~38$ calculated at level four is shown in Fig.~\ref{convergence} as
a function
of the dimensionality, both for $c_n=0$ and $c_n=-1$. For lower dimensions
the former family of solutions is much better. Still, even this gives a
value larger than 0.25, which one can obtain with a pair of real qubits only if $n\geq 12$.
Unfortunately, this family does not converge to the upper bound. The
other family with $c_n=-1$ is initially inferior, but above $n=79$ it
overtakes the other one, and converges to $0.250~875~384~514$, a value
well consistent with the upper bound: it is slightly larger, but within
the numerical accuracy of the bound. In Fig.~\ref{aszimcl} and Fig.~\ref{szimcl} we show $\lambda_i$ and
$c_i$ for $n=99$ with $c_n=0$ and $c_n=-1$, respectively. For small $i$ the $c_i$
is close to $c_0$=0, and tends to a constant value fast. In the case of
$c_n=0$ it stays basically constant up to the last few point, when it drops
towards the zero value of $c_n$ fast. When $c_n=-1$ and $n>19$ there are more
than one solutions. In all solutions $c_i>0$ for small $i$ and $c_i<0$ for large $i$.
In the different solutions the change of sign happens between two
neighbouring integers. For very large $n$ it always happens above $16$ and below $n-16$.
Among all these solutions the best is the one in which $c_i$ changes sign exactly in
the middle, that is between $(n-1)/2$ and $(n+1)/2$. In Fig.~\ref{szimcl} we show the result for
this solution. In particular, the upper curve in Fig.~\ref{szimcl} shows the behaviour of $c_i$. 
The curve starts like in
the case of $c_n=0$, that is it drops from $c_0=1$ to a constant value fast,
that stays there until it goes to minus one times the constant value in a relatively
short interval, and then it stays almost constant again until it turns towards $c_n=-1$
at the end. Where $c_i$ is constant, $\lambda_i$ behaves exponentially.

Knowing the behaviour of $c_i$, it is easy to choose initial values such that the
iterative procedure will converge to the desired solution for the first time,
there is no need for repeated runs. For example, we may initialize $c_i$ as
$+C$ for $i<n/2$ and  $-C$ for $i>n/2$, with $C=0.9$. Then we can start with
the third step instead of the first one, that is with the determination of $\lambda_i$.

The families of solutions may be extended to even values of $n$. The matrices of
the measurement operators are blockdiagonal, like for odd $n$. The matrices
of $\hat B_1$, $\hat B_2$ and $\hat A_3$ consist of $n/2$ two by two blocks. The blocks
of $\hat A_1$, $\hat A_2$ and $\hat B_3$ are also two-dimensional except for the
first and the last ones, which have single elements. The actual form of the matrices
can be parametrized with $c_i$ analogously to the odd $n$ case, which
will satisfy the same equations as optimality conditions, and the eigenvalue equation
for $\lambda_i$ will also have the same form. Figure~\ref{convergence} does contain the results
for $n$ even. The curve for $c_i$ in the case of the best solution with $c_n=-1$ is not
fully symmetric. It changes either very nearly half way between $n/2-1$
and $n/2$ or between $n/2$ and $n/2+1$. If we make the change of size to happen at
the middle, that is we enforce $c_{n/2}=0$, for large $n$ the result will converge
to the same suboptimal value as in the case of the $c_n=0$ family. Probably the reason
this family fails to converge to the true optimum is that $c_i$ must approach zero at an
integer value, instead of midway between two integers.

In the matrices of the measurement operators we made very specific
choices for the values of the single element blocks, leaving only $c_n$ as a single free
parameter. We note that there are other choices consistent with the optimality
conditions, but they lead to solutions equivalent with the ones discussed.  

\section{Other examples}\label{other}

\begin{table*}[t]
\caption{Results for the $19$ Bell inequalities for which the tightness of
the quantum bound has not been proven. The local Hilbert space necessary to achieve the best
lower bound is denoted by $\cal H$, column 'New' tells if this lower bound is an improvement
over our previous result, and column 'Level' is the level to derive the upper bound.}
\vskip 0.2truecm
\centering
\begin{tabular}{l c l l c l}
\hline\hline

Case\hphantom{W}&Lower bound&$\cal H$&New&Upper bound-&\hphantom{ww}Level\\
&&&&lower bound&\\
 \hline
$I^{20}_{4422}$&0.4676794&$\RR^{4}$&no&0.0000000&$L3+aa'bb'$\\
$J^{30}_{4422}$&0.4363842&$\RR^{17}$&yes&0.0008787&$L3$\\
$J^{34}_{4422}$&0.4492657&$\RR^{4}$&no&0.0000000&$L3+aa'bb'$\\
$J^{48}_{4422}$&0.7516220&$\RR^{12}$&yes&0.0000120&$L3+aa'bb'$\\
$J^{62}_{4422}$&0.7500755&$\RR^{6}$&no&0.0000166&$L3$\\
$J^{73}_{4422}$&1.1223170&$\RR^{20}$&yes&0.0002409&$L3$\\
$J^{97}_{4422}$&1.1583626&$\RR^{2}$&no&0.0001425&$L3$\\
$A_{14}$&0.4759513&$\RR^{20}$&yes&0.0018953&$L3$\\
$A_{21}$&0.3260601&$\RR^{10}$&yes&0.0001176&$L3$\\
$A_{47}$&0.4608544&$\CC^{2}$&no&0.0020847&$L2+aa'b+abb'$\\
$A_{62}$&0.4065268&$\RR^{15}$&yes&0.0003457&$L2+aa'b+abb'$\\
$A_{64}$&0.3900890&$\RR^{3}$&no&0.0013878&$L2+aa'b+abb'$\\
$A_{65}$&0.3688996&$\RR^{5}$&yes&0.0000000&$L2+aa'b+abb$\\
$A_{67}$&0.3990671&$\RR^{5}$&no&0.0000000&$L3$\\
$A_{68}$&0.4025522&$\RR^{18}$&yes&0.0024807&$L2+aa'b+abb'$\\
$A_{80}$&0.3769863&$\RR^{4}$&no&0.0002557&$L2+aa'b+abb'$\\
$A_{82}$&0.4708838&$\RR^{20}$&yes&0.0000341&$L2+aa'b+abb'$\\
$A_{84}$&0.6352087&$\RR^{18}$&yes&0.0000747&$L2+aa'b+abb'$\\
$A_{89}$&0.3035637&$\RR^{17}$&yes&0.0022575&$L2+aa'b+abb'$\\
 \hline
 \end{tabular}
 \label{table:highlowdiff}
 \end{table*}

In Ref.~\cite{PV09} we calculated both lower and upper bounds for all known tight binary bipartite Bell
inequalities with up to five measurement settings per party. For most cases we managed to
find explicit solutions saturating the upper bound. There were 20 exceptions. The case of
$I_{3322}$ was discussed in the previous chapter. Now we have applied the iterative algorithm
with up to 25 dimensional component spaces for the remaining 19 problems.
In 11 cases we have found better solution than the ones we reported in Ref.~\cite{PV09}. The solution for $A_{65}$ could have been found by our previous methods, but we missed it. All the other ones involve larger Hilbert spaces than those methods could handle. We have also
made the upper bound tighter by going to higher levels by using a computer with larger
memory. As previously, the calculations have been done with Borcher's code CSDP for
semidefinite programming~\cite{CSDP}.
For cases with five measurement settings per party we could do the calculation on level
two plus $aa'b$ plus $abb'$. For the smaller cases we could afford to go up to level
three, except for $I_{3322}$, where we did level four. The notion of levels and partial levels,
and their notation is explained in Ref.~\cite{NPA2}, and also in Ref~\cite{PV09}. This way the two bounds met for $A_{65}$. We sent $A_{67}$ at level three, and $J^{48}_{4422}$ and $J^{97}_{4422}$ at
level three plus $aa'bb'$ to Brian Borchers, as test cases for his new code which requires
much less memory than CSDP for very large cases. For $A_{67}$ this new upper bound does
agree with the lower one. Unfortunately, for $J^{48}_{4422}$ there remained a difference of
$0.000012$, significantly less than the level three value of $0.0000657$, but more than numerical
uncertainty. At an even higher level it would probably disappear, but it is not sure. Therefore,
we can not state that our solution with 12-dimensional component spaces is already the optimum one.
The case $J^{97}_{4422}$ at level three plus $aa'bb'$ seems to be a difficult case for the new
code, it could solve it only at a reduced accuracy, and the value it has given for the
upper bound turned out to be lower than the lower bound. Therefore, very probably the
solution with a pair of real qubits we found the best so far is actually the optimum.
If we added to level three just one quarter of the $aa'bb'$ type terms, we could use CSDP.
In the cases of $I^{20}_{4422}$ and $J^{34}_{4422}$ this was already enough to reach
the lower bound. By looking at the dependence of the upper bound on the level it has been
calculated, we are almost sure that for $J^{30}_{4422}$, $J^{62}_{4422}$, $A_{47}$,
$A_{62}$, $A_{64}$, $A_{82}$ and $A_{89}$ the present lower bound will not be reached,
therefore we do not yet have the optimum solution. Probably $I_{3322}$  is not the only
inequality requiring infinite dimensional Hilbert spaces for maximum violation. Actually, the
difference in the cases of $J^{62}_{4422}$ and $A_{82}$ is very small, but it does not
change much with increasing level. For $A_{80}$ and $A_{84}$ the upper bound would more
probably converge to our solution, while for the rest of the cases we can not tell from the
present results.

Bancal et al.~\cite{BGP} found recently two symmetric inequalities with four settings per party,
which they called $S_{51}$ and $S_{52}$. They were also included in our calculations. In fact, both
of them could be saturated with measurements acting on qubits,
we found the maximum violation to be $1.0135274$ and $0.87038004$ for $S_{51}$ and $S_{52}$,
respectively. 

The details of our best solutions, including Schmidt coefficients and
the matrices of the measurement operators are on our web site~\cite{Bellweb}.

\section{Conclusion}\label{conc}

To summarize, we investigated the $I_{3322}$ Bell inequality and its maximal quantum violation. Although $I_{3322}$ is one of the simplest Bell inequalities with just three measurements per party, surprisingly we found that the maximum value of 0.25 achievable with a pair of qubits could be only be overcome by using states of dimension at least $12\times 12$. Moreover, we found a family of measurement operators and states, which attains numerically in very large dimensional Hilbert spaces (and hence in the limit of infinite dimensions as well) the largest possible value of the Bell expression $I_{3322}$ allowed by quantum theory. Taking as a conjecture that our construction is optimal (in the sense that no finite dimension suffices to achieve the true quantum maximum for $I_{3322}$), to our knowledge this constitutes the first example of a Bell scenario concerning finite number of measurements and outcomes, where measuring finite dimensional quantum systems is not enough to obtain exactly all quantum correlations.  In this respect we wish to invoke the concept of dimension witnesses introduced recently in Ref.~\cite{Brunner}. In a broader sense, the aim of this concept is to put a lower bound on the Hilbert space dimension needed to reproduce the statistics arising from a Bell experiment. Dimension witnesses have been found or conjectured in many different scenarios, such as for the bipartite setting \cite{PV08,PV09,bounddim,Junge,VPB,Zohren} for the multipartite setting \cite{Perez-Garcia} and even for the case of a single system \cite{single}.

In particular, for measurements with binary outcomes it was shown how to get dimension witnesses analytically for any dimensions \cite{bounddim}. Actually, these results entail that no finite dimension is sufficient to generate the whole set of quantum correlations provided an arbitrary number of two-outcome measurements is involved. Consider now a scenario where the number of measurements (each having a finite number of outcomes) is finite. Does there exist a Bell inequality requiring infinitely large entangled states for obtaining the maximum possible quantum violation? That is, are there quantum correlations corresponding to this scheme attainable exactly by measuring infinite dimensional quantum systems? This is a question posed recently by Navascu\'es et al.~\cite{NPA2}.

Indeed, in the present paper we suggested, supported by thorough numerical computations, that no finite dimension suffices to maximally violate the $I_{3322}$ inequality. While we cannot guarantee that for any finite dimension there exist no other construction beyond ours giving a larger quantum value, nevertheless we showed that our best conjectured values tend to the upper limit computed with the aid of the NPA method \cite{NPA1}. We pose it as a challenge to provide an analytical proof of our conjecture. Also, in our view it would be interesting to find other, bipartite or multipartite, Bell inequalities beyond $I_{3322}$ (but still involving finite number of settings and outcomes) requiring infinitely large entangled states for their maximal quantum violation.

\acknowledgments T.V. has been supported by a J\'anos Bolyai Grant of the Hungarian Academy of
Sciences. We thank for the help of Brian Borchers, who used some of our problems as test cases
for his code for semidefinite programming, and this way provided us with improved upper bounds
for those cases. We would like to thank Miguel Navascu\'es for illuminating discussions.

\end{document}